\documentstyle[iopconf1,epsfig,psfig]{article}
\newcommand{\newc}{\newcommand} 
\newc{\ra}{\rightarrow} 
\newc{\lra}{\leftrightarrow} 
\newc{\beq}{\begin{equation}} 
\newc{\eeq}{\end{equation}} 
\newc{\barr}{\begin{eqnarray}} 
\newc{\earr}{\end{eqnarray}} 
\begin{document}
\title{Double Beta Decay in Gauge Theories}

\author{J.D. Vergados \footnote{ Permanent address:
Theoretical Physics Section, University of Ioannina,
GR--451 10, Ioannina, Greece\\
E-mail:vergados@cc.uoi.gr}
}
\affil{ 
 Institute of Theoretical Physics, University of Tuebingen,
D--72076 Tuebingen, Germany
}

\date{\today}

\beginabstract
Neutrinoless double beta decay is a very important process 
both from the particle and nuclear physics point of view. 
From the elementary particle point of view it pops up in almost
every model. In addition to the traditional mechanisms,
like the light neutrino mass,
$\lambda$ and $\eta$ terms etc we can have direct R-parity violating
supersymmetric (SUSY) contributions. In any case its observation  
will severely constrain the existing models and 
will signal that the neutrinos are massive Majorana particles.
From the nuclear
physics point of view it is challenging, because: 1) The relevant nuclei
have complicated nuclear structure. 2) The
energetically allowed transitions are exhaust a small part of all
the strength. 3) One must
cope with the short distance behavior of the transition operators, especially
when the intermediate particles are heavy (eg in SUSY models). Thus novel 
effects, like  the double beta decay of pions in flight between nucleons, 
have to be considered.
4) The intermediate momenta involved are about 100 $MeV/c$. Thus 
one has to take into account  possible momentum 
dependent terms in the nucleon current.
We find that, for the mass mechanism, such modifications of the
nucleon current for light neutrinos reduce the nuclear matrix elements by about
$25\%$, almost regardless of the nuclear model.
In the case of heavy neutrino the effect is much larger and model dependent.
 Taking the above effects into account, the  availabe nuclear 
matrix elements for the experimentally  interesting nuclei A = 76,
82, 96, 100, 116, 128, 130, 136 and 150 and the presently available
experimental limits on the half-life of the $0\nu\beta\beta$-decay we have 
extracted the following new limits: 
$\langle m_{\nu} \rangle < 0.3eV/c^2$ 
$\lambda^{\prime}_{111} < 4.0 \times 10^{-4}$
for the R-parity violating parameter with 
reasonable choices of the parameters of SUSY models,
\endabstract
\section{Introduction}

The nuclear double beta decay can occur whenever the ordinary (single) beta
decay is forbidden due to energy conservation or greatly suppressed due to
angular momentum mismatch. The exotic neutrinoless double beta decay 
($0\nu\beta\beta-decay$) is the most interesting since it violates lepton number
by two units. It is a very old process. It was first considered by Furry 
\cite {Fur39} exactly half a century ago as soon it was realized that the
neutrino might be a Majorana particle. It was continued with the work of
Primakoff and Rosen \cite{PR52} especially when it
was recognized that kinematically it is favored by $10^{8}$ compared to its non
exotic sister $2\nu\beta\beta$-decay.  When the corresponding level of the
$10^{15y}$ lifetime was reached and the process was not seen, it was tempting
to interpret this as an indication that the neutrino was a Dirac particle.
The interest in it was resurrected with the advent of gauge theories which favor
Majorana neutrinos and through the pioneering work of Kotani and his group
\cite {DTNOT} it was brought again to the attention of the nuclear physics 
community. To-day, fifty years later, $0\nu\beta\beta$-decay continues to be
one of the most interesting processes.

 From a theoretical point of view it is the most
likely, if not the only, process capable of deciding whether or not the
neutrino is a Majorana particle, i.e. it coincides with its own antiparticle
\cite{HS84,DTK85,Ver86,Tom91,SC98,FS98}.
It is expected to occur whenever one has
lepton number violating interactions. Lepton number, being a global quantity
is not sacred, but it is expected to be broken at some level. In short this
process pops up almost everywhere, in every theory.

From a nuclear physics
point of view calculating the relevant nuclear matrix elements it is indeed
a challenge. First almost all nuclei, which can undergo double beta decay,
are far from closed shells and some of them are even deformed. One thus faces
a formidable task. Second the nuclear matrix elements represent a small
fraction of a canonical value ( i.e. the matrix element to the energy non
allowed transition double Gamow-Teller resonance or some appropriate sum
rule). Thus effects which are normally negligible become important here.
Third in many models the dominant mechanism for $0\nu\beta\beta$-decay
does not involve intermediate light neutrinos, but very heavy particles
and one must be able to cope with the short distance behavior of the
relevant operators and wave functions.

From the experimental point of view it also very challenging to measure
the slowest perhaps process accessible to observation. Especially since it
is realized that, even if one obtains only lower bounds on the life
time for this decay, the extracted limits  on the theoretical model
parameters may be comparable, if not better, and complementary to those
extracted from the most ambitious accelerator experiments.

The recent superkamiokande results have given the first evidence of physics
beyond the Standard Model (SM)  and in particular they indicate that the
neutrinos are massive
particles. It is important to proceed further and find out whether the
neutrinos are Dirac or Majorana particles. As we have mentioned there might
be processes other than the conventional intermediate neutrino mechanism,
which may dominate $0\nu\beta\beta$-decay. It has, however, been known that
whatever the lepton violating process is, which gives rise to this decay,
it can be used to generate a Majorana mass for the neutrino \cite{SVa82}. 
The study of the $0\nu\beta\beta$-decay is further stimulated by the 
development of grand unified theories (GUT's) and supersymmetric
models (SUSY) 
representing extensions of the $SU(2)_L \otimes U(1)$ SM. 
The GUT's and SUSY offer a variety of mechanisms
which allow the $0\nu\beta\beta$-decay to occur \cite{Moh98}.

 The best known 
possibility is via the exchange of a Majorana neutrino between the two
decaying neutrons \cite{HS84,DTK85,Ver86,Tom91,PSV96,FS98,SC98}. 
 Nuclear physics dictates that we study the 
light and heavy neutrino components separately. In the presence of only
left-handed currents for light intermediate neutrinos the obtained amplitude
is proportional to a suitable average neutrino mass, which vanishes in the limit in which the neutrinos become Dirac particles.
In the case of heavy Majorana neutrino components the amplitude is proportional
to the average inverse neutrino mass, i.e. it is again suppressed.  In the
 presence of right handed currents one has one can have a contribution similar
 to the one above for heavy neutrinos but involving a different (larger)
average inverse  mass and some suppression due to the heaviness of $W_R$. 

It is also possible to have, in addition,
interference between the leptonic left and right currents, $j_{L}-j_{R}$ 
interference. In this case the amplitude in momentum space becomes proportional
to the 4-momentum of the neutrino and, as a result, only the light neutrino
components become important. One now has two possibilities. First the two
hadronic currents have a chirality structure of the same kind $J_L-J_R$.
Then one can extract from the data a
dimensionless parameter $\lambda$, which is proportional to the square of
the ratio of the masses of the L and R gauge bosons, $\kappa=(m_{L}/m_{R})^2$.
Second the two hadronic currents are left-handed, which can happen via the
mixing of the two bosons. The relevant lepton violating parameter $\eta$ is
now proportional to this mixing. Both of these parameters, however, involve
the neutrino mixing and they are proportional to the mixing between the 
light and heavy neutrinos.

 In gauge theories one has, of course, many more possibilities. Exotic 
intermediate scalars may mediate $0\nu\beta\beta$-decay \cite{Ver86}. These are
not favored in current gauge theories and are not going to be further
discussed. In superstring inspired models one may have siglet fermions in
addition to the usual right handed neutrinos. Not much progress has been made
on the phenomenological side of these models and they are not going to be
discussed further.

 In recent years supersymmetric models are taken seriously and semirealistic
calculations are taking place. In standard calculations one invokes
universality at the GUT scale, employing in all 5 parameters, and use the
renormalization group equation to obtain all parameters (couplings and
particle masses) at low energies.  Hence, since such parameters are in
principle calculable one can use $0\nu\beta\beta$-decay to constrain some
of the R-parity violating couplings, which cannot be specified by the theory 
\cite{Moh86,Ver87,HKK95,WKS97,FKSS97,FKS98a,FKS98b}. 
Recent review articles \cite{FS98,SC98} give  a detailed account of the
latest developments in this field. 

 From the above discussion it clear that one has to consider the case of
heavy intermediate particles. One thus has to consider very short ranged
operators in the presence of the nuclear repulsive core. If the interacting
nucleons are point-like one gets negligible contributions. We know, however
that the nucleons are not point like and they have structure described by
a form factor, which can be calculated in the quark model or parameterized
in a dipole shape. This approach was first considered by Vergados \cite{Ver81}
adopted later by almost everybody. The resulting effective operator has
a range somewhat less than the proton mass (see sect. 4 below).
 
The other approach is to consider particles other than the nucleons present
in the nuclear soup. For $0^+\rightarrow 0^+$ the most important such particles
are the pions. One thus may consider the double beta decay of pions in flight
between nucleons, like
\beq
                   \pi^- \longrightarrow \pi^+~e^-~e^-~~~~~  ,~~~~~~
                   n \longrightarrow p~\pi^+~e^-~e^- 
\label{eq:1}   
\eeq
 This contribution was first considered by Vergados \cite{Ver82} and was found
to yield results of the same order as the nucleon mode with the above recipe
for treating the short range behavior. It was revived by the Tuebingen group 
\cite{FKSS97,FKS98a} in the context of R-parity violating interactions, in
which it appears to dominate.

 The other recent development is the better description of nucleon current
by including momentum dependent terms, such as the modification of the axial
current due to PCAC and the inclusion of the weak magnetism terms. These
contributions have been considered previously \cite{tom85,PSV96}, but only
in connection with
the extraction of the $\eta$ parameter mentioned above. Indeed these terms
were very important in this case since they compete with the p-wave lepton wave
function, which, with the usual currents, provides the lowest non vanishing
contribution. In the mass
term, however, only s-lepton wave functions are relevant. So these terms have
hitherto  been neglected.

 It was recently found \cite {SPVF} that for light
neutrinos the inclusion of these momentum dependent terms reduces the nuclear
matrix element by about $25\%$, independently of the nuclear model employed.
For the heavy neutrino, however, the effect can be larger and depends on the
nuclear wave functions.
 The reason for expecting them to be relevant is that the average momentum
$<q>$ of the exchanged neutrino is expected to be large \cite{SEIL92}. In the
case of a light intermediate neutrino the  mean nucleon-nucleon separation is
about 2 fm which implies that the average momentum $<q>$ is about 100 MeV.  
In the case of a heavy neutrino
exchange the mean internucleon distance is considerably smaller 
and the average momentum $<q>$ is supposed to be considerably larger. 

 Since $0\nu-\beta \beta$ decay is a two step process, one should in principle
construct and sum over all the intermediate nuclear steps, a formidable job
indeed in the case of the Shell Model Calculations (SMC). Since, however, the
average neutrino momentum is much larger compared to the nuclear excitations,
one can invoke closure using some average excitation energy (this does not
apply in the case of $2\nu \beta \beta$ decays). Thus one need construct only
the initial and final $0^+$ nuclear states. In Quasiparticle Random Phase
Approximation  (QRPA) one must construct the intermediate states anyway. In any
case it was explicitly shown, taking advantage of the momentum space formalism
developed by Vergados \cite {Ver90}, that this approximation is very good
 \cite {PV90,FKPV91}. The same conclusion was reached independently by others
 \cite{SKF90}.

Granted that one takes into account all the above ingredients in order to 
obtain quantitative answers for
the lepton number violating parameters from the the results of
$0\nu\beta\beta$-decay experiments, it is necessary to evaluate the relevant
nuclear matrix elements with high reliability. The most extensively use
methods  are the  SMC ( for a recent review see \cite
{SC98}) and QRPA( for a recent 
review see \cite {FS98,SC98}). The SMC is forced to use few single particle
 orbitals,
while this restriction does not apply in the case of QRPA. The latter suffers
, of course, from the approximations inherent in the RPA method. So a direct
comparison between them is not possible.

 The SMC has a long history 
\cite{Ver76,HSS82,SV83,ZBR90,CPZ90,SSDV92,ZB93} in in double beta decay
calculations. In recent years it has lead to large matrices calculations  in
traditional as well as Monte Carlo types of calculations
\cite{RCN95,Retal96,CNPR96,NSM96,SDSJ97,KDL97} (For a more complete set of
references see Ref. \cite{SC98}) and suitable effective interactions.

 There have been a number of QRPA calculations covering almost all nuclear
targets 
\cite{VZ86,CAT87,MBK88,EVJP91,RFSK91,GV92,SC93,CS94,SSVP97,SPF98,cheoun}.
 We also have seen some refinements of QRPA, like proton neutron pairing and
inclusion of 
renormalization effects due to Pauli principle corrections \cite{TS95,SSF96}. 

 The above schemes, in conjunction with the other
improvements mentioned above offer, some optimism in our efforts for obtaining
nuclear matrix
elements  accurate enough to allow us to extract reliable values of the lepton
violating parameters from the data.  We will review this in the case
of most of the nuclear targets of experimental
interest ($^{76}{{Ge}}$, $^{82}{{Se}}$, $^{96}{{Zr}}$, $^{100}{{Mo}}$,
$^{116}{{Cd}}$, $^{128}{{Te}}$, $^{130}{{Te}}$, $^{136}{{Xe}}$,
$^{150}{{Nd}}$).

\section{Theory}

\subsection{Majorana neutrino mass mechanism}

We shall consider the $0\nu\beta\beta$-decay process assuming 
that the  effective beta decay Hamiltonian acquires the form:
\begin{equation}
{\cal H}^\beta = \frac{G_{{F}}}{\sqrt{2}} 
\left[\bar{e} \gamma_\mu (1-\gamma_5) \nu^{0}_{{e L}} \right]
J^{\mu \dagger}_L + 
\left[\bar{e} \gamma_\mu (1+\gamma_5) \nu^{0}_{{e R}} \right]
J^{\mu \dagger}_R + {h.c.}
\label{eq:1.1}   
\end{equation}
where $e $ and $\nu^{0}_{{e L}}$ $\nu^{0}_{{e R}}$ are field operators
representing the electron and the left handed and the right handed electron
neutrinos in the weak interaction basis, respectively. 
We suppose that  neutrino mixing does take place according to
\begin{equation}
\nu^{0}_{e L}=\sum^3_{k=1} ~U^{(11)}_{ek}~\nu_{kL} +
\sum^3_{k=1} ~U^{(12)}_{ek}~N_{kL},
\label{eq:1.2}   
\end{equation}
\begin{equation}
\nu^{0}_{e R}=\sum^3_{k=1} ~U^{(21)}_{ek}~\nu_{kL} +
\sum^3_{k=1} ~U^{(22)}_{ek}~N_{kL},
\label{eq:1.3}   
\end{equation}
where, $\nu_{k}$ ($N_{k}$)
are fields of light (heavy) Majorana neutrinos with masses
$m_k$ ($m_k << 1$ MeV) and $M_k$ ($M_k >> 1$ GeV), respectively. 
The matrices  $U^{(11)}_{ek}$  and $U^{(22)}_{ek}$ are approximately unitary,
while the matrices 
$U^{(12)}_{ek}$  and $U^{(21)}_{ek}$ are very small (of order of the up
quark  divided by the heavy neutrino mass scales) so that the overall matrix 
is unitary. 
$\nu_k,N_k$ satisfy the Majorana condition: 
$\nu_k \xi_k = C ~{\overline{\nu}}_k^T$, 
$N_k \Xi_k = C ~{\overline{N}}_k^T$,
where C denotes the charge conjugation and $\xi$, $ \Xi$ 
are phase factors (the eigenmasses are assumed positive).

We assume both outgoing electrons to be
in the $s_{1/2}$ state and consider only 
$0^+_i\rightarrow 0^+_f$ transitions are allowed. 
For the ground state transition restricting ourselves to the
mass mechanism we obtain for the
$0\nu\beta\beta$-decay inverse half-life 
\cite{HS84,DTK85,Ver86,Tom91,PSV96,FS98,SC98},
\begin{equation}
[T_{1/2}^{0\nu}]^{-1} = G_{01} 
[|\frac{<m_\nu >}{m_e}M^{light}_{<m_\nu >} + 
\eta^{L}_{_{N}} M^{heavy}_{\eta_{_N}}|^2 +
|\eta^{R}_{_N} M^{heavy}_{\eta_{_N}}|^2]
\label{eq:1.4}   
\end{equation}
The lepton-number non-conserving parameters, i.e. the
effective neutrino mass $<m_\nu >$ and 
$\eta^{L}_{_N}$ ,$\eta^{R}_{_N}$  are given as follows:
\begin{eqnarray}
<m_\nu > ~ = ~ \sum^{3}_1~ (U^{(11)}_{ek})^2 ~ \xi_k ~ m_k, ~ ~~~~~~~
\eta^L_{_N} ~ = ~ \sum^{3}_1~ (U^{(12)}_{ek})^2 ~ 
~\Xi_k ~ \frac{m_p}{M_k},
\label{eq:1.5}   
\end{eqnarray}
\beq
\eta^R_{_N} ~ = ~ ( \kappa ^2 + \epsilon ^2) \sum^{3}_1~ (U^{22}_{ek})^2 ~ 
~\Xi_k ~ \frac{m_p}{M_k},
\label{eq:1.6}   
\eeq
with $m_p$ ($m_e$) being the proton (electron) mass, $\kappa$ 
is the mass squared ratio of $W_L$ and $W_R$ and $\epsilon$ their mixing. 
$G_{01}$ is the integrated kinematical factor \cite{DTK85,PSV96}. 
The nuclear matrix elements associated with the
exchange of light ($M^{light}_{<m_\nu >}$) and heavy neutrino
($M^{heavy}_{\eta_{_N}}$) must be computed in a nuclear model.
Eq. (\ref{eq:1.4}), however, applies to any intermediate particle. 

 At this point we should stress that the main suppression in the mass terms
comes from the smallness of neutrino masses. In the case of heavy neutrino
not only from the large values of neutrino masses but the small couplings,
$U^{(12)}$ for the left handed neutrinos and $\kappa$ and $\epsilon$ for the
right-handed ones.

\subsection{The leptonic left-right interference mechanism ($\lambda$ and
$\eta$ terms).}

 As we have already mentioned in the presence of right handed currents one
can have interference between the leptonic currents of opposite chirality.
This leads to different kinematical functions and two new lepton violating
parameters $\lambda$ and $\eta$ defined by
\beq
\eta ~ = ~ \epsilon ~\eta_{RL}~~~,~~~ \lambda ~=~ \kappa ~\eta_{RL}~~~,~~~ 
   \eta_{RL} ~=~ \sum^{3}_1~ (U^{(21)}_{ek}U^{(11)}_{ek}) ~\xi_k 
\label{eq:1.7}   
\eeq
 The parameters $\lambda$ and $\eta$ are small not only due to the smallness
of the parameters $\kappa$ and $\epsilon$ but in addition
because of the smallness of $U_{(21)}$. 

All the above contributions vanish in
the limit in which the neutrino is a Dirac particle.

 Many nuclear matrix elements appear in this case, but they are fairly well
known and they are not going to be reviewed here (see e.g. \cite{HS84,Ver86}
and \cite {DTK85,Tom91,SC98,FS98} and in our notation \cite {PSV96}). We only
mention that in the case of the $\eta$ we have additional contributions coming
from the nucleon recoil term and the kinematically favored spin antisymmetric
term. These  dominate and lead to values of $\eta$ much smaller than $\lambda$
\cite {PSV96}.

\section{The R-parity violating contribution}
 In SUSY theories R-parity is defined as
\beq
R = (-1)^{3 B + L + 2 s}
\label{eq:2.1}   
\eeq
with $B=baryon,L=lepton$ numbers and s the spin. It is +1 for ordinary
particles and -1 for their superpartners.  R-parity violation has recently
been seriously considered in SUSY models. 

R-parity violating terms may induce 
a Majorana neutrino mass and may, therefore, lead to $0\nu \beta \beta$ decay. But the relevant masses are small, at least 10 times smaller 
\cite{HVFK99}
than those deduced from the present data. The bilinear terms in the
 superpotential also lead to mixings between the neutrinos and neutralinos as 
well as between the leptons and the charginos,
leading to lepton violating processes \cite{HV99}. We are not going, however,
to consider such effects in this review.

 Here we will be concerned with trilinear couplings in the superpotential
given by:
\beq
W= \lambda_{ijk} L^a_i L^b_j E^c_k \epsilon_{ab} +
              \lambda^{\prime}_{ijk} L^a_i U^b_j D^c_k \epsilon_{ab} +
              \lambda^{\prime\prime}_{ijk} U^c_i U^c_j D^c_k
\label{eq:2.2}   
\eeq
where a summation over the flavor indices i,j,k and the isospin indices
a,b is understood ( $\lambda_{ijk}$ is antisymmetric in the indices i and j)
The last term has no bearing in our discussion, but we will assume that it 
vanishes due to some discreet symmetry to avoid too fast proton decay.
The $\lambda$'s are dimensionless couplings not predicted by the theory.

In the above notation L,Q are isodoublet and $E^c,D^c$ isosinglet chiral
superfields, i.e they represent both the fermion and the scalar components.
 It has been recognized quite sometime ago that the second term in the
superpotential could lead to neutrinoless double beta decay \cite {Moh86,Ver87}
and re-examined quite recently \cite {FKSS97}. Typical diagrams at the quark
level are shown in Fig.1. Note that as intermediate states, in
addition to the s-leptons and s-quarks, one must consider the neutralinos,
4 states which are linear combinations of the gauginos and higgsinos, and
the colored gluinos (supersymmetric partners of the gluons). 
\begin{figure}[ht]
\epsfxsize=12cm
\epsfysize=10cm
\begin{center}
\epsffile{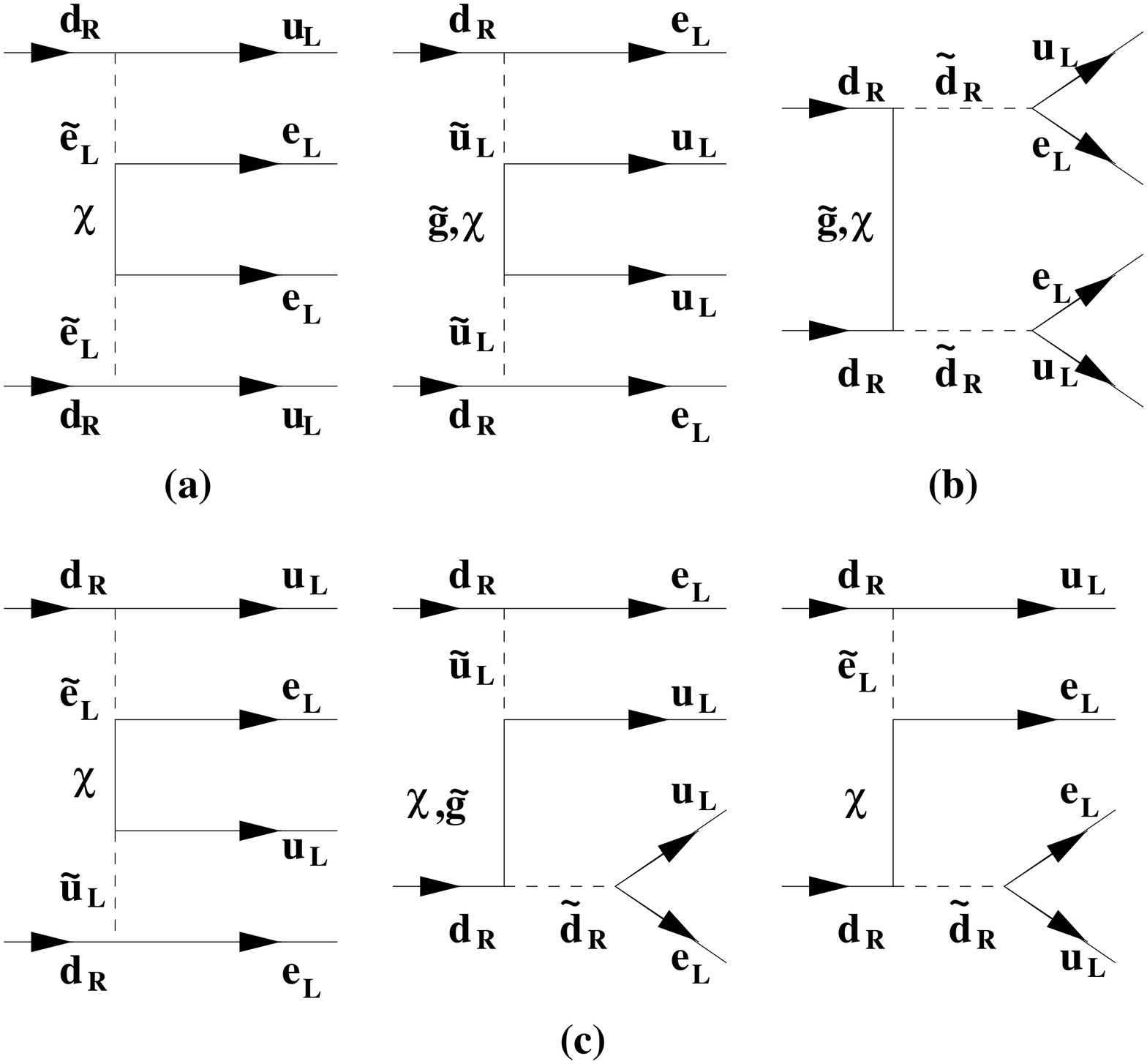}
\caption[]{
The R-parity violating contribution to $0\nu \beta \beta$ decay mediated by
s-fermions and neutralinos (gluinos).}
\label{Fig.1}
\end{center}
\end{figure}
 Whenever the process is mediated by gluons a Fierz transformation is
needed to lead to a colorless combination. The same thing is necessary
whenever the fermion line connects a quark to a lepton. As a result
one gets at the quark level not only scalar (S) and pseudoscalar (P) couplings,
but tensor (T) couplings as well. This must be contrasted to the V and A 
structure of the traditional mechanisms. One, therefore, must consider
how to transform these operators from the quark to the nucleon level.

The effective lepton violating parameter, assuming that pion exchange mode
dominates, as the authors of Ref \cite {FKSS97,FS98} claim, is given by
\beq
\eta_{SUSY} = (\lambda^{\prime}_{111})^2 \frac{3}{8}(\chi_{PS}~ \eta_{PS}+
 \eta_T)
\label{eq:2.3}   
\eeq
with $\eta_{PS}(\eta_T)$ associated with the scalar and pseudoscalar (tensor)
quark couplings given by
\beq
\eta_{PS} = \eta_{\tilde{\chi},\tilde{e}}+ \eta_{\tilde{\chi},\tilde{q}}+
 \eta_{\tilde{\chi},\tilde{f}}+
               \tilde{\eta}_{\tilde{g}}+ 7 \eta^{\prime}_{\tilde{g}} ~~,~~
\eta_{T} =  \eta_{\tilde{\chi},\tilde{q}}- \eta_{\tilde{\chi},\tilde{f}}+
              \tilde{\eta}_{\tilde{g}} - \eta^{\prime}_{\tilde{g}}
\label{eq:2.5}   
\eeq
 They find $\chi_{PS}=(2/3)$, but as we shall see it depends on ratios of 
nuclear matrix elements.
For the diagram of Fig.1a one finds
\beq
\eta_{\tilde{\chi},\tilde{e}} = \frac{2\pi \alpha}{(G_Fm_W^2)^2} 
             (\kappa_{\tilde{e}})^2 \langle\frac{m_p}{m_{\tilde{\chi}}}
             \rangle_{\tilde{e}\tilde{e}}        
\label{eq:2.6}   
\eeq
For the diagram of Fig.1b one finds
\beq
 \tilde{\eta}_{\tilde{\chi},\tilde{q}} = \frac{\pi \alpha}{2(G_Fm_W^2)^2} 
             [(\kappa_{\tilde{d}})^2 \langle\frac{m_p}{m_{\tilde{\chi}}}
             \rangle_{\tilde{d}\tilde{d}} +        
             (\kappa_{\tilde{u}})^2 \langle\frac{m_p}{m_{\tilde{\chi}}}
             \rangle_{\tilde{u}\tilde{u}}]       
\label{eq:2.7}   
\eeq
\beq
 \tilde{\eta}_{\tilde{g}} = \frac{\pi}{6} \alpha _s \frac{1}{(G_F m_W^2)^2} 
   [(\kappa_{\tilde{d}})^2 +(\kappa_{\tilde{u}})^2] \frac{m_p}{m_{\tilde{g}}}
\label{eq:2.8}   
\eeq
For the diagram of Fig.1c one finds
\beq
 \tilde{\eta}_{\tilde{\chi},\tilde{f}} =  \frac{\pi \alpha}{2(G_Fm_W^2)^2} 
  [\kappa_{\tilde{e}} \kappa_{\tilde{d}} \langle\frac{m_p}{m_{\tilde{\chi}}}
             \rangle_{\tilde{e}\tilde{d}}  +      
  \kappa_{\tilde{e}} \kappa_{\tilde{u}} \langle\frac{m_p}{m_{\tilde{\chi}}}
             \rangle_{\tilde{e}\tilde{u}}  +      
  \kappa_{\tilde{d}} \kappa_{\tilde{u}} \langle\frac{m_p}{m_{\tilde{\chi}}}
             \rangle_{\tilde{d}\tilde{u}}]      
\label{eq:2.9}   
\eeq 
\beq
 \tilde{\eta}_{\tilde{g}^{\prime}} = \frac{\pi}{12} \alpha _s 
\frac{1}{(G_F m_W^2)^2} 
   \kappa_{\tilde{d}}\kappa_{\tilde{u}} \frac{m_p}{m_{\tilde{g}}}
\label{eq:2.10}   
\eeq 
where
\beq 
  \kappa_X = (\frac{m_W}{m_X})^2 \, , \, X=\tilde{e}_L,\tilde{u}_L~~~~~,~~~~~
  \kappa_{\tilde{d}} = (\frac{m_W}{m_{\tilde{d}_R}})^2 
\label{eq:2.12}   
\eeq 
\beq 
\langle\frac{m_p}{m_{\tilde{\chi}}} \rangle_{\tilde{f}\tilde{f}^{\prime}} =  
                      ~\sum_1^4~ \epsilon_{\tilde{\chi}_{i},\tilde{f}} 
\epsilon_{\tilde{\chi}_{i},\tilde{f}^{\prime}} \frac{m_p}{m_{\tilde{\chi}_{i}}}
\label{eq:2.13}   
\eeq
where $\epsilon_{\tilde{\chi}_{i},\tilde{f}}$ and 
$\epsilon_{\tilde{\chi}_{i},\tilde{f}^{\prime}}$ are the couplings of the 
$i^{th}$ neutralino to the relevant fermion-sfermion, which are calculable
( see e.g Ref. \cite{Ver96}). Thus ignoring the small Yukawa couplings coming
via the Higgsinos and taking into account only the gauge couplings we find
\beq 
\epsilon_{\tilde{\chi}_{i},\tilde{e}} = \frac{Z_{2i}+tan \theta _W Z_{1i}}
                                              {sin \theta _W} 
\label{eq:2.14}   
\eeq 

\beq 
\epsilon_{\tilde{\chi}_{i},\tilde{u}} = \frac{Z_{2i}+(tan \theta_W ~/3) Z_{1i}}
                                              {sin \theta _W} ~~~,~~~
\epsilon_{\tilde{\chi}_{i},\tilde{d}} = -\frac{Z_{1i}}
                                              {3 cos \theta _W} 
\label{eq:2.16}   
\eeq 
where $Z_{1i},Z_{2i}$ are the coefficients in the expansion of the $\tilde{B},
\tilde{W}_{3}$ in terms of the neutralino mass eigenstates. Note that in this
convention some of the masses $m_{\tilde{\chi}_{i}}$  may be negative.
\section{The effective nucleon current}
 As we have mentioned the effective nucleon current in addition to the
usual V and A terms (P,S,T in SUSY contributions) contains momentum
dependent terms  \cite{SPVF}. 

Within the impulse approximation the nuclear current  $J^\rho_L$
in Eq. (\ref{eq:1})  expressed with nucleon fields $\Psi$ takes the form
\begin{eqnarray}
J^{\mu \dagger}_L 
=  \overline{\Psi} \tau^+ \left[ g_V(q^2) \gamma^\mu 
- i g_M (q^2) \frac{\sigma^{\mu \nu}}{2 m_p} q_\nu
 - g_A(q^2) \gamma^\mu\gamma_5 + g_P(q^2) q^\mu \gamma_5 \right] \Psi,
\label{eq:3.1}   
\end{eqnarray}
where M is the nucleon mass, $q^\mu = (p-p')_\mu$ is the momentum transferred
from hadrons to leptons ($p$ and $p'$ are four momenta of neutron and 
proton, respectively) and 
$\sigma^{\mu\nu} = (i/2)[\gamma^{\mu}, \gamma^{\nu}]$.
$g_V(q^2)$, $g_M(q^2)$, $g_A(q^2)$ and $g_P(q^2)$ are real functions
of a Lorenz scalar $q^2$. The values of these form factors in the
zero-momentum transfer limit are known as the vector, weak-magnetism,
axial vector and induced pseudoscalar coupling constants, respectively. 

For nuclear structure calculations it is necessary to reduce the nucleon
current to the non-relativistic form. We shall neglect small energy transfers
between nucleons in the non-relativistic expansion. Then the form of the
nucleon current coincides with those in the Breit frame  and we 
arrive at \cite{ERI88},
\begin{equation}
J^\mu(\vec{x})=\sum_{n=1}^A \tau^+_n [g^{\mu 0} J^0({\vec q}^{~2}) +
g^{\mu k} J^k_n({\vec q}^{~2})] \delta(\vec{x}-{\vec{r}}_n),~~~~k=1,2,3,
\label{eq:3.2}   
\end{equation}
with 
\begin{eqnarray}
J^0({\vec q}^{~2}) = g_V(q^2),~~
{\vec J}_n({\vec q}^{~2}) = g_M({\vec q}^{~2}) 
i \frac{{\vec{\sigma}}_n \times \vec{q}}{2 M}+
g_A({\vec q}^{~2}) [\vec{\sigma}
-\frac{\vec{q}~ {\vec{\sigma}}_n \cdot \vec{q}}{ \vec{q}^2+m^2_{\pi}}]
\label{eq:3.3}   
\end{eqnarray}
${\vec r}_n$ is the coordinate of the $n$th nucleon.

For the form factors we shall use the following 
parameterization \cite {SPVF}: 

$g_V({\vec q}^{~2}) = {g_V}/{(1+{\vec q}^{~2}/{\Lambda^2_V})^2}$, 
$g_M({\vec q}^{~2})= (\mu_p-\mu_n) g_V({\vec q}^{~2})$, 

$g_A({\vec q}^{~2}) = {g_A}/{(1+{\vec q}^{~2}/{\Lambda^2_A})^2}$  

where $g_V = 1$, $g_A = 1.254$, $(\mu_p-\mu_n) = 3.70$, 
$\Lambda^2_V = 0.71 ~(GeV)^2$ \cite{dumb} and $\Lambda_A = 1.09 ~GeV$ 
\cite{towner}. In previous calculations only one 
general cut-off $\Lambda_V = \Lambda_A \approx 0.85$ GeV
was used. In this work 
we take the empirical value of $\Lambda_A $  deduced from the
antineutrino quasielastic reaction 
${\overline{\nu}}_\mu p \rightarrow \mu^+ n $. A larger value of
the cut-off $\Lambda_A $ is expected to increase slightly the values of
corresponding nuclear matrix elements.  
 It worth noting that with these modifications of the nuclear current
one gets a new contribution in the neutrino mass mechanism, namely the
tensor contribution.
The two body effective transition operator takes in momentum space the
form
\beq
\Omega = \tau_+ \tau_+ (- g_V({\vec q}^{~2}) + h_{GT}~ \sigma_{12} - h_T S_{12})
\label{eq:3.4}   
\eeq
where the three terms  correspond to Fermi (F) , Gamow-Teller (GT) and Tensor
(T).  One finds that
\beq
S_{12} = 3({\vec{ \sigma}}_1\cdot \hat{q}
       {\vec{\sigma}}_2 \cdot \hat{q})
      - \sigma_{12}, ~~~ \sigma_{12}=
{\vec{ \sigma}}_1\cdot {\vec{ \sigma}}_2.
\label{eq:3.5}   
\eeq
Note that the tensor operator is defined in momentum space ( $\hat{q}$
rather than $\hat{r}$) and there is a change of sign in going to the
coordinate space.
\begin{eqnarray}
\frac{h_{GT} ({\vec q}^{~2})}{g^2_A({\vec q}^{~2})} & = &~[ ~1~ -~ 
\frac{2}{3}~ \frac{ {\vec q}^{~2}}{ {\vec q}^{~2} + m^2_\pi } ~+ ~
\frac{1}{3} ~( \frac{ {\vec q}^{~2}}{ {\vec q}^{~2} + m^2_\pi } )^2 ~]
~ +~ \frac{2}{3}~ \frac{g^2_M ({\vec q}^{~2} ) {\vec q}^{~2} }{4 m^2_p }, 
\nonumber \\
\frac{h_T ({\vec q}^{~2})}{ g^2_A({\vec q}^{~2})~} & = & [~ 
\frac{2}{3}~ \frac{ {\vec q}^{~2}}{ {\vec q}^{~2} + m^2_\pi } -
\frac{1}{3}~ ( \frac{ {\vec q}^{~2}}{ {\vec q}^{~2} + m^2_\pi } )^2 ~]~
~+~ \frac{1}{3} ~\frac{g^2_M ({\vec q}^{~2} ) {\vec q}^{~2} }{4 m^2_p },
\label{eq:3.6}    
\end{eqnarray}

\begin{table}[t]
\footnotesize\rm
\caption{The Fermi, Gamow-Teller and Tensor 
nuclear matrix elements for the light Majorana neutrino
exchange of the $0\nu\beta\beta$-decay of
$^{76}Ge$ and $^{130}Te$ with ( rows 2 and 4) and without (rows 1 and3)
short-range correlations.}
\label{table.1}
\begin{tabular}{lrrrrrrrr}
\topline
transition  &\multicolumn{3}{c}{ Gamow-Teller} & 
  \multicolumn{2}{c}{Tensor} & $M^{light}_F$ & 
  $M^{light}_{GT}$ & $M^{light}_T$ \\
    & AA & AP & PP & AP & PP & & & \\ \hline
$^{76}Ge $  & 
 5.132 & -1.392 & 0.302 & -0.243 & 0.054 & -2.059 & 4.042  & -0.188 \\
& 2.797 & -0.790 & 0.176 & -0.246 & 0.055 & -1.261 & 2.183 & -0.190 \\
\hline
$^{130}Te $ &
 4.158 & -1.173 & 0.258 & -0.329 & 0.074 & -1.837 & 3.243 & -0.255 \\
 &1.841 & -0.578 & 0.134 & -0.333 & 0.075 & -1.033 & 1.397 & -0.258 \\
\bottomline
\end{tabular}
\end{table}

\begin{table}[t]
\footnotesize\rm
\caption{Nuclear matrix elements for the light and heavy 
Majorana neutrino exchange modes of the $0\nu\beta\beta$-decay for the
nuclei studied in this work calculated within
the renormalized pn-QRPA. }
\label{table.2}
\begin{tabular}{lrrrrrrrrr}
\topline
\multicolumn{10}{c}{ $(\beta\beta)_{0\nu}-decay: 
0^{+}\rightarrow{0^{+}}$ 
 transition} \\ \cline{2-10}
M. E. & $^{76}Ge$ & $^{82}Se$ & $^{96}Zr$ & $^{100}Mo$ &
 $^{116}Cd$ & $^{128}Te$ & $^{130}Te$ & $^{136}Xe$ & $^{150}Nd$ 
\\ \hline
\multicolumn{10}{c}{ light Majorana neutrino  (I=light)} \\
$M_{VV}^I$ &
 0.80  &   0.74  &   0.45  &   0.82  &  0.50  &
 0.75  &   0.66  &   0.32  &   1.14 \\
$M_{AA}^I$ &
 2.80  &   2.66  &   1.54  &   3.30  &  2.08  &
 2.21  &   1.84  &   0.70  &   3.37 \\
$M_{PP}^I$ &
 0.23  &   0.22  &   0.15  &   0.26  &  0.15  &
 0.24  &   0.21  &   0.11  &   0.35 \\
$M_{AP}^I$ &
-1.04  &  -0.98  &  -0.65  &  -1.17  &  -0.69 &
-1.04  &  -0.91  &  -0.48  &  -1.53 \\
 & & & & & & & & & \\
$M_{<m_\nu >}^I$ &
 2.80  &   2.64  &   1.49  &   3.21  &   2.05 &
 2.17  &   1.80  &   0.66  &   3.33 \\
 & & & & & & & & & \\
\multicolumn{10}{c}{ heavy Majorana neutrino  (I= heavy)} \\
$M_{VV}^I$ &
 23.9  &   22.0  &   16.1  &   28.3  &   17.2 &
 25.8  &   23.4  &   13.9  &   39.4 \\
$M_{MM}^I$ &
-55.4  &  -51.6  &  -38.1  &  -67.3  &  -39.8 &
-60.4  &  -54.5  &  -31.3  &  -92.0 \\
$M_{AA}^I$ &
 106.  &   98.3  &   68.4  &   123.  &   74.0 &
 111.  &   100.  &   58.3  &   167. \\
$M_{PP}^I$ &
 13.0  &   12.0  &   9.3   &   16.1  &   9.1  &
 14.9  &   13.6  &   7.9   &   23.0 \\
$M_{AP}^I$ &
-55.1  &  -50.7  &  -41.1  &  -70.1  &  -39.0 &
-64.9  &  -59.4  &  -34.8  &  -101. \\
 & & & & & & & & & \\
$M_{\eta_{_N}}^I$ &
 32.6  &  30.0   &   14.7  &   29.7  &   21.5 &
 26.6  &  23.1   &   14.1  &   35.6 \\
\bottomline
\end{tabular}
\end{table}
 The exact results will depend on the details of the nuclear model, since
the new operators have different momentum (radial) dependence than the
traditional ones and the tensor component is entirely new. We can get a crude
idea of what is happening by taking the above average momentum
 $\langle q \rangle$=100 MeV/c. Then we find that the GT ME is reduced by
22$\%$. Then assuming that T matrix element is about half the GT one, we
find that the total reduction is 28$\%$. This is in perfect agreement with
the exact results for the A=76 system, $29\%$, but a bit smaller than the $38\%$
obtained for the A=130 system.  
 We will now summarize the results obtained with the 
 above modifications of the nucleon current for light neutrino 
(see Eq. (\ref{eq:1.4})) for the two representative $0\nu\beta\beta$-decay 
nuclei 
$^{76}Ge$ and $^{130}Te$ in Table \ref{table.1}.
The details of our calculations will be given elsewhere \cite{SPVF}
One notices significant additional contributions to GT (AP and PP)
and tensor (AA and PP) nuclear matrix elements coming from
higher order nucleon current terms. AP and PP originate from the second (first)
and third (second) terms in $h_{GT}(h_T)$ of Eq. (\ref{eq:3.6}).

 By glancing at the 
Table \ref{table.1} we also see that, with proper treatment of
short-range two-nucleon correlations (see e.g Vergados \cite {Ver86}), all 
matrix elements are strongly suppressed. The effect is even stronger in the
case of heavy intermediate particles. Detailed results \cite {SPVF} for
various nuclei are presented in Table \ref{table.2}.

\section{Extraction of the lepton violating parameters}

The limits deduced for the lepton-number violating parameters depend 
on the values of nuclear matrix element, of the kinematical factor and
of the current experimental limit for a given isotope
[see Eq. (\ref{eq:1.4})].

\subsection{Traditional lepton violating parameters}

Even, though, we expect the nuclear matrix elements entering the light neutrino
mass mechanism to be decreased by about $30\%$ , independently of the nuclear
model, we will stick to the calculations as reported. Thus the present best
experimental limits \cite{hdmo97}$^-$\cite {moe94} can be converted to upper
limits on  $<m_\nu >$ and $\eta_{_N}$. 
\begin{table}[t]
\footnotesize\rm
\caption{The present state of the
Majorana neutrino mass searches in
$\beta\beta$-decay experiments.
$T^{exp-0\nu}_{1/2}$(present) 
is the best presently available lower limit on
the half-life of the $0\nu\beta\beta$-decay   
for a given isotope.
The corresponding upper limits on lepton number non-conserving parameters 
${<m_\nu >}$  and $\eta_{N}$  are presented. 
For the definition of the references and "best" see text.}
\label{table.3}
\begin{tabular}{llllll}
\topline
Nucleus  & $^{76}Ge$ & $^{82}Se$ & $^{96}Zr$ & $^{100}Mo$ & $^{116}Cd$ \\
\hline
 $T^{exp-0\nu}_{1/2}$(present) [y] &
 $ 1.1\times 10^{25}$    & $  2.7\times 10^{22}$ &
 $ 3.9\times 10^{19}$    & $  5.2\times 10^{22}$ &  $ 2.9\times 10^{22}$ \\
Ref. & [Ex1] & [Ex2] & [Ex3] & [Ex4] & [Ex5] \\ 
 $<m_\nu >$ [eV]  & 0.62 & 6.3 & 203. & 2.9 & 5.9 \\
 $T^{exp-0\nu}_{1/2} [y]$ & & & & &\\ 
 $(<m_\nu >^{best})$  &  
 $1.1\times 10^{25}$ & $2.8\times 10^{24}$ & $4.2\times 10^{24}$ &
 $1.2\times 10^{24}$ & $2.6\times 10^{24}$ \\
 $\eta_{_N} $ & $1.0\times 10^{-7}$ & $1.1\times 10^{-6}$ & 
 $4.0\times 10^{-5}$ &
 $6.2\times 10^{-7}$ & $1.1\times 10^{-6}$ \\   
 $T^{exp-0\nu}_{1/2} (\eta_{_N}^{best})$ [y] &  
 $1.1\times 10^{25}$ & $2.9\times 10^{24}$ & $5.8\times 10^{24}$ &
 $1.8\times 10^{24}$ & $3.2\times 10^{24}$ \\
 & & & & & \\
\hline
Nucleus & $^{128}Te$ & $^{130}Te$ & $^{136}Xe$ & $^{150}Nd$ & \\
\hline
 $T^{exp-0\nu}_{1/2}$(present) [y] &
 $  7.7\times 10^{24}$   &  $  8.2\times 10^{21}$  &
 $  4.2\times 10^{23}$   &  $  1.2\times 10^{21}$  &  \\
Ref.  & [Ex6] & [Ex7] & [Ex8] & [Ex9] & \\
 $<m_\nu >$ [eV]  & 1.8 & 13. & 4.9 & 8.5  \\
 $T^{exp-0\nu}_{1/2}$  & & & & &\\ 
 $ (<m_\nu >^{best})$ [y] & $6.6\times 10^{25}$ & $3.8\times 10^{24}$ &
 $2.7\times 10^{25}$ & $2.3\times 10^{23}$ \\
 $\eta_N $ & $2.9\times 10^{-7}$ & $2.0\times 10^{-6}$ & $4.5\times 10^{-7}$ &
 $1.6\times 10^{-6}$ \\   
 $T^{exp-0\nu}_{1/2} [y] (\eta_{_N}^{best})$ [y] &  
 $5.9\times 10^{25}$ & $3.1\times 10^{24}$ & $7.9\times 10^{24}$ &
 $2.7\times 10^{23}$ \\
\bottomline
\end{tabular}
\end{table}

\begin{table}[t]
\footnotesize\rm
\caption{The lifetimes  predicted for $0^+ \rightarrow 0^+$ 
$0\nu\beta\beta$-decay in various mechanisms (light neutrino, heavy neutrino,
$\lambda$ and $\eta$ terms and SUSY contribution) for suitable input of lepton
violating parameters and available nuclear calculations.For the definitions of
the references see text.}
\label{table.4}
\begin{tabular}{lrrrrrrrrrr}
\topline
\multicolumn{11}{c}{ $(\beta\beta)_{0\nu}-decay: 
0^{+}\rightarrow{0^{+}}$ transition} \\ 
\multicolumn{11}{c}{ $T^{0\nu -theor}_{1/2}(\langle m_{\nu}\rangle,
 \langle \lambda \rangle, \langle \eta \rangle, \langle \eta _N\rangle,
 \langle \eta _{SUSY} \rangle)$} \\ 
\cline{2-11}
&$^{48}Ca$ & $^{76}Ge$ & $^{82}Se$ & $^{96}Zr$ & $^{100}Mo$ &
 $^{116}Cd$ & $^{128}Te$ & $^{130}Te$ & $^{136}Xe$ & $^{150}Nd$ \\
\cline{2-11}
Ref. &$10^{24}$ & $10^{24}$ & $10^{24}$ & $10^{24}$ & $10^{24}$ & $10^{24}$ &
$10^{25}$ & $10^{24}$ & $10^{24}$ & $10^{22}$ \\ 
\hline
\multicolumn{10}{c}{$\langle m_{\nu} \rangle=1eV,\langle \lambda \rangle=0,
\langle \eta \rangle=0,\langle \eta _N\rangle=0,\langle \eta _{SUSY}\rangle=0$}
\\ 
R & 12.8 & 34.8 & 4.80 &  &  &  & & & 24.2&\\
H & 6.34 & 3.36 & 1.16 &  &  &  & 0.80 & 0.32 & &\\
E1 &  & 4.60 & 1.84 &  &  &  & 0.90 & 0.48 & &\\
E2 &  & 28.0 & 11.2 &  &  &  & 3.00 & 1.32 & 6.60 &\\
S &  & 8.12 & 2.86 &  &  &  & 3.60 & 1.66 &  &\\
M &  & 4.66 & 1.20 &  & 2.54  &  & 1.54 & 0.98 & 4.42  & 6.74\\
T &  & 4.32 & 1.22 &  & 0.52  &  & 1.96 & 1.08 & 2.80  & 8.90\\
P1 & 5.00  & 7.20 & 3.00 & 1.22 & 7.80  & 9.40  & 3.80 & 1.72 & 6.60  & \\
P2 & 56.0  & 36.0 & 5.60 & 54.0 &  & 9.80  & 30.0 & 4.20 & 5.60  & \\
S1 &  & 17.9 &  &  & 0.50  & 1.44 & 2.18 &   & 17.5 & \\
P &  & 4.22 & 1.08 & 1.61 & 0.46  & 0.99 & 2.53 & 1.46  & 10.1 & 8.78\\
\multicolumn{10}{c}{$\langle m_{\nu} \rangle=0,\langle \lambda \rangle=10^{-6},
\langle \eta \rangle=0,\langle \eta _N\rangle=0,\langle \eta _{SUSY}\rangle=0$} \\ 
R & 7.45 & 50.2 & 3.25 &  &  &  & & & 22.2&\\
S &  & 7.75 & 1.14 &  &  &  & 14.8 & 0.89 &  &\\
M &  & 7.35 & 0.99 &  & 0.95  &  & 13.5 & 0.95 & 4.90  & 3.73\\
T &  & 8.02 & 1.07 &  & 0.55  &  & 21.1 & 1.18 & 3.47  & 6.71\\
P1 & 2.71  & 8.90 & 2.08 & 0.94 & 30.6 & 39.1  & 22.7 & 1.34 & 2.73  & \\
P2 & 27.9 & 41.2 & 4.39 & 27.7 & 10.3& 10.8 & 165 & 2.22  & 4.42 &\\
\bottomline
\end{tabular}
\end{table}
\begin{table}[t]
\footnotesize\rm
\caption{The previous table continued.} 
\label{table.5}
\begin{tabular}{lrrrrrrrrrr}
\topline
\multicolumn{10}{c}{$\langle m_{\nu} \rangle=0,\langle \lambda \rangle =0,
\langle \eta \rangle=10^{-8},\langle \eta _N\rangle=0,
\langle \eta _{SUSY}\rangle=0$} \\ 
R & 6.42 & 27.2 & 6.24 &  &  &  & & & 22.2&\\
S &  & 36.7 & 11.1 &  &  &  & 10.7 & 5.92 &  &\\
M &  & 7.35 & 0.99 &  & 0.95  &  & 13.5 & 0.95 & 4.90  & 3.73\\
T &  & 2.25 & 0.65 &  & 0.28 &  & 0.67 & 0.44 & 1.21  & 3.39\\
P1 & 15.11  & 3.10 & 6.51 & 1.48 & 3.44 & 19.2  & 1.20 & 0.62 & 1.23  & \\
P2 & 43.2 & 22.8 & 5.16 & 7.95 & 102 & 83.2 & 1.90  & 1.05 & 0.96 &\\
\multicolumn{10}{c}{$\langle m_{\nu} \rangle=0,\langle \lambda \rangle=0,
\langle \eta \rangle=0,\langle \eta _N\rangle=10^{-7},
\langle \eta _{SUSY}\rangle=0$} \\ 
P1 & 4.95  & 0.25 & 3.35 & 67.1 & 4.70 & 23.5  & 0.78 & 3.03 & 1.42  & \\
P2 & 124 & 0.59 & 7.23 & 671 & 1.47 & 33.6 & 1.27  & 1.31 & 1.01 &\\
P & & 15.4  & 4.10 & 8.10 & 0.97 & 8.40 & 8.51 & 4.54 & 3.94 &40.6 \\
\multicolumn{10}{c}{$\langle m_{\nu} \rangle=0,\langle \lambda \rangle=0,
\langle \eta \rangle=0,\langle\ eta _N\rangle =0,
\langle \eta _{SUSY} \rangle =10^{-8}$} \\ 
F  & & 3.3 & 0.86 & 0.71 & 0.30 & 0.85 & 0.93 & 0.45 & 1.2 & 3.2 \\
P & & 4.5 & 1.0 & 1.4 & 0.59 & 1.4 & 1.5 & 0.71 & 2.0 & 5.2 \\
\bottomline
\end{tabular}
\end{table}

The thus obtained results are given in Table \ref{table.4}. The references of 
Table \ref{table.4} are defined as follows:
Ex1=Heidelberg-Moscow Collaboration \cite{hdmo97},
Ex2=Elliott {\it et al} \cite{ell92},
Ex3=Kawashima {\it et al} \cite{kaw93},
Ex4=Ejiri {\it et al} \cite{eji96}, 
Ex5=Davenich {\it et al} \cite{dane95}, 
Ex6=Bernatovicz {\it et al} \cite{bern92},
Ex7=Alessandrello {\it et al} \cite{ale94},
Ex8=De Silva {\it et al} \cite{bus96},
Ex9=Busto {\it et al} \cite{sil97} 
Thus, the most restrictive limits are as follows:
\beq
<m_\nu >^{best} ~< ~~ 0.62~ eV ~~~~,~~~~ 
<\eta_{_N} >^{best}~< ~~1.0\times 10^{-7}
\label{vio.2}
\eeq
(see Ref.\cite{SPVF}$,$ \cite{hdmo97})
By  assuming  $<m_\nu > = <m_\nu >^{best}$ and   
$\eta_{_N} = \eta_{_N}^{best}$.  
(\ref{eq:1.4})  
we calculated half-lives of the $0\nu\beta\beta$-decay
$T^{exp-0\nu}_{1/2}$($<m_\nu >^{best}$), 
$T^{exp-0\nu}_{1/2}$($\eta^{best}_{_N}$)  
for nuclear systems of interest
using specific mechanisms with the "best" parameters. 
The thus obtained results are given in Table \ref{table.4}. 
The references of 
Table \ref{table.4} are defined as follows:
Ex1=Heidelberg-Moscow Collaboration \cite{hdmo97},
Ex2=Elliott {\it et al} \cite{ell92},
Ex3=Kawashima {\it et al} \cite{kaw93},
Ex4=Ejiri {\it et al} \cite{eji96}, 
Ex5=Davenich {\it et al} \cite{dane95}, 
Ex6=Bernatovicz {\it et al} \cite{bern92},
Ex7=Alessandrello {\it et al} \cite{ale94},
Ex8=Busto {\it et al} \cite{sil97}.
Ex9=De Silva {\it et al} \cite{bus96},
Since the quantities $<m_\nu >$, $\eta_{_N}$
depend only on particle theory parameters these quantities 
indicate the experimental half-life limit for a
given isotope, which the relevant experiments should 
reach in order to extract the best
present bound on the corresponding lepton number violating 
parameter from their data. Some of them have a long way to go to
reach the $Ge$ target limit.

 A summary involving most of the available nuclear matrix elements and taking
into account what, at present, is a good guess as canonical values of the 
lepton violating parameters 
is provided in Tables \ref{table.4} and  \ref{table.5}. 
The references in these tables are
defined as follows: R=Retamosa {\it et al} \cite{RCN95},
 H=Haxton {\it et al} \cite{HSS82}, E1=Engel   {\it et al} \cite{EVJP91},
 E2=Engel   {\it et al} \cite{VZ86}, S=Suhonen  {\it et al} \cite{SKF90},
 M=Muto   {\it et al} \cite{MBK88}, T=Tomoda {\it et al} \cite{Tom91},
 P1=Pantis {\it et al} \cite{PSV96}, P2=Pantis {\it et al} \cite{PSV96}
(p-n pairing), 
 S1=Simkovic {\it et al} \cite{SPF98} (and private communication),
 F=Faessler {\it et al} \cite{FKS98a}$,$ \cite {FKS98b}
 P=Present calculation (see Simkovic {\it et al} \cite{SPVF} for the
nuclear Matrix elements). 
Notice in particular that the present
calculation, marked P in the table, involves not only renormalized QRPA
 \cite {SEIL92,SSF96}, but takes into account the corrections in the hadronic
current \cite {SPVF} discussed above (see table \ref{table.4}).

  
\subsection{R-parity induced lepton violating parameters}


 In this section we will elaborate a bit further on the R-parity violating
parameters. We will consider the pionic contribution (\ref{eq:1}). We will
first attempt to evaluate the relevant amplitude using harmonic oscillator
wave functions, but adjusting the parameters to fit related experiments.

 Let us begin with the second process of Eq. (\ref{eq:1}). This process involves
a direct term and an exchange term. 
The direct term is nothing but a decay
of the pion into two leptons with a simultaneous 
change of a neutron to a proton by the relevant nucleon current,  which
in this case can only be of the of the PS type. The tensor contribution
cannot lead to a pseudoscalar coupling at the nucleon level, which is
needed to be coupled to the usual pion nucleon coupling in the other vertex
to get the relevant operator for a $0^+~ \rightarrow~0^+$ decay. Thus the
amplitude involving the meson is related to $\pi,\mu$ decay.
\beq
 A_{1 \pi}(direct) = 4~\tilde{\alpha }_{1\pi}  
      \frac{\sigma_1.\vec{q}}{(2~m_N)} m^2_{\pi} ~~exp(-\frac{(qb)^2}{6} 
\label{eq:6.a}   
\eeq
with
\beq
\tilde{\alpha}_{1\pi}~=~(2 \pi)^3~ \frac{m_N}{3~m_q}~ \chi (0)
\label{eq:6.c}   
\eeq

The exchange contribution, in which the produced up quark of the meson is not
produced from the $"vacuum"$ but it comes from the initial nucleon, is a bit 
more complicated. The harmonic oscillator quark model, however can be used to
get its relative magnitude (including the sign) with respect to the direct term.
 This way we find 
\beq
 A_{1 \pi}(exchange) = -3~\tilde{\alpha }_{1\pi}  
      [\frac{\sigma_1.\vec{q}}{(2~m_N)} m^2_{\pi}
  ~exp(-\frac{(qb)^2}{6}] 
\label{eq:6.b}   
\eeq
Thus the effective two-body
transition operator in momentum space at the nucleon level becomes:
\beq
\Omega^{PS}_{1\pi}~=~ c_{1\pi}~~
      [\frac{\sigma_1.\vec{q}\sigma_2.\vec{q}}{(2~m_N)^2}
  ~exp(-\frac{(qb)^2}{6}] \frac{m_{\pi}^2}{q^2+m_{\pi}^2}
\label{eq:6.1}   
\eeq
with $c_{1\pi}=~g_r~ \tilde{\alpha}_{1\pi}$ i.e.
\beq
c_{1\pi}~=~(2 \pi)^3~ \frac{m_N}{3~m_q}~ g_r~\chi (0)
\label{eq:6.d}   
\eeq

where $g_{r}=13.5$ is the pion nucleon coupling and $m_q$ is
the constituent quark mass. Note the presence of the exponential form factor
in the harmonic oscillator model, which has been ignored in other treatments.  
We see  that, in going from the quark to the nucleon
level, the factor of three coming from the mass
gain is lost due to the momentum being reduced by a factor of three. The  
quantity $\chi (0)$ is essentially the meson wave function at the origin
given by:
\beq
  \chi (0) = \frac{\sqrt{6}}{2^{1/4}}~m_{\pi}^{-3/2}~\psi (0) 
\label{eq:6.2a}   
\eeq
 The quantity $\chi (0)$ can be obtained from the $\pi \rightarrow \mu , \nu$
decay via the expression
\beq
  \frac{1}{\tau} = \frac{1}{\pi} (\frac{G_F}{\sqrt{2}})^2~m_{\pi}^2~m_{\mu}^2
                  (1-\frac{m^2_{\pi}}{m^2_{\mu}})^2~\chi ^2(0)
\label{eq:6.2b}   
\eeq
From the measured lifetime $\tau =2.6 \times 10^{-8}$ we obtain 
$\chi (0)=0.46$

 The first process of Eq. \ref{eq:1} is easier to handle. Now both the PS and
T terms contribute. We thus get
\beq
A_{2 \pi}(T) = \frac{3}{8} ~\tilde{\alpha}_{2\pi}~m^4~,
A_{2 \pi}(PS) = \frac{1}{8} ~\tilde{\alpha}_{2\pi}~m^4~,
\label{eq:6.5a}   
\eeq 
\beq
\tilde{\alpha}_{2\pi}~=~4~(2 \pi)^3~\chi^2(0)
\label{eq:6.5b}   
\eeq 
\beq
\Omega^{T}_{2\pi}~=~ c_{2 \pi}
     ~\frac{\sigma_1.\vec{q}\sigma_2.\vec{q}}{(2~m_N)^2}
 \frac{m_{\pi}^4}{(q^2+m_{\pi}^2)^2} ~~,~~
\Omega^{PS}_{2\pi}~=~ \frac{2}{3} \Omega^{T}_{2\pi}
\label{eq:6.4}   
\eeq
\beq
\tilde{\alpha}_{2\pi}~=~4~(2 \pi)^3~\chi^2(0) ~~~,~~~
c_{2\pi}~=~4~(2 \pi)^3~g^2_r ~\chi ^2(0)
\label{eq:6.5}   
\eeq 
 Using the above value of $\chi(0)$ and $g_r=13.5$ we get 
$c_{1\pi}~=~109$~ and $c_{2\pi}~=~198$~
which are in good agreement with the values $132.4$ and $170.3$ respectively
obtained by Faessler et al \cite{FKS98a}.

   It is now customary, but it can be avoided \cite{Ver90}, to go to coordinate
space and express the nuclear
matrix elements in the same scale with the standard matrix elements involving 
only nucleons. Thus we get
\beq
ME_k~ = (\frac{m_A}{m_p})^2~ \alpha_{k\pi}~\frac{m_p}{m_e}~
        [ M^{k\pi}_{GT}+M^{k\pi}_T]~
\label{eq:6.6}   
\eeq
Where the two above matrix elements are the usual GT and T matrix elements
with the additional radial dependence given by
\beq
F^{1\pi}_{GT} ~ = e^{-x}~~~  ,~~~ F^{1\pi}_T ~ = (3+3x+x^2)~\frac{e^{-x}}{x}~
\label{eq:6.7}   
\eeq
\beq
F^{2\pi}_{GT} ~ = (x-2)e^{-x}~~~  ,~~~ F^{2\pi}_T ~ = (1+x)~e^{-x}~
\label{eq:6.8}   
\eeq

\beq
\alpha_{1\pi}~=~-~ c_{1\pi}~\rho~, 
               \alpha_{2\pi}~=~ c_{2\pi}~ \rho ~~~,~~~
    \rho = \frac{1}{48 f^2_A} ~(\frac{m_{\pi}}{m_p})^4 (\frac{m_p}{m_A})^2
\label{eq:6.10}   
\eeq
 In the above formulas we have tried to stick to the
definition of  $\eta_{SUSY}$ given above (see \ref {eq:2.3}), but since the
tensor (at the quark level) does not contribute to the $1\pi$ diagram the
"effective" nuclear matrix element is not the sum of the two matrix elements
of Eq. (\ref{eq:6.6}), but only $ME_2$, and $\chi_{PS}$ depends on the nuclear
matrix elements, i.e.
\beq
ME_{eff} = ME_2 ~~~~~,~~~~\chi_{PS}~ = ~ \frac{2}{3}~ (4\frac{ME_1}{ME_2} + 1) 
\label{eq:6.11}   
\eeq
 There is no difference, of course, between the two expressions if $ME_2$ is
dominant, as is actually the case).

 Before proceeding further we should remark that for the experimentally
derived harmonic oscillator parameter for the $\pi $ meson, $b=1.8~f$, 
$\alpha_{2\pi}$ s dominant and $\chi_{PS}$ approaches the value of 2/3. 
 In fact  we find 
$\alpha_{1\pi}~=~-1.2 \times 10^{-2}$~ and $\tilde{\alpha}_{1\pi}~=~0.15$
which are in good agreement with the values $-4.4 \times 10^{-2}$ and $0.20$
respectively obtained by Faessler et al \cite{FKS98a}.
Furthermore from the nuclear
matrix elements of Ref. \cite{FKS98a} one can see that the $M^{2\pi}$  is
favored, since, among other things, its tensor and Gamow-Teller components 
are  the same magnitude and sign (in the $1\pi$ mode they are
opposite). Thus nuclear physics also favors the $2\pi$ mode.

With the above ingredients and using the nuclear matrix elements of 
\cite{FKS98a} we can extract from the data values of $\eta_{SUSY}$.

Then one can use these
values of $\eta_{SUSY}$ in order to extract values for the R-parity violating
parameters $\lambda^{\prime}_{111}$.

 As we have already mentioned one must start with 5 parameters in the allowed
SUSY parameter space and solve the RGE equations to obtain the values of the
needed parameters at low energies \cite{KANE}$^-$ \cite{WKS99}. For our 
purposes is adequate to utilize
typical parameters, which have already appeared in the literature 
\cite{KANE}$^,$ \cite{CPR94}.
One then finds
\beq
\lambda^{\prime}_{111}=C_{\tilde{\chi} ^0}(\eta_{SUSY})^{1/2}~~~(neutralinos~
 only)
\label{eq:6.12}   
\eeq
\beq
\lambda^{\prime}_{111}=C_{\tilde{g}}(\eta_{SUSY})^{1/2} ~~~~(gluino~only)
\label{eq:6.13}   
\eeq
When both neutralinos and gluinos are included we write
\beq
\lambda^{\prime}_{111}=C_{\tilde{\chi} ^0, \tilde{g}}(\eta_{SUSY})^{1/2}
\label{eq:6.14}   
\eeq
 The values of these coefficients are given in Tab. 6 for the nine SUSY
models mentioned above.
\begin{table}[t]
\footnotesize\rm
\caption{A sample of relevant parameters obtained by some choices in the allowed
SUSY parameter space. It clear that in all cases the neutralino mediated
mechanism is dominant (for definitions see text). The parameters C shown have
been multiplied by $10^{-3}$} 
\label{table.6}
\begin{tabular}{lrrrrrrrrr}
\topline
input &\multicolumn{3}{c}{ Kane et al (\# 1-3)} & 
  \multicolumn{6}{c}{Ramond et al (\# 4-9)} \\ 
\hline
 & \# 1 & \#2 & \#3 &\#4  &\#5  &\#6  &\#7 &\#8 &\#9\\
\hline
$tan\beta$ & 10. & 1.5 & 5.0 &5.4  &2.7  &2.7  & 5.2 & 2.6 & 6.3\\
$m_{\chi ^0 _1} $ & 124. &  26 &  96 & 83 &124 &58  & 34 & 34 & 50\\
$m_{\chi ^0 _2} $ & 237. &  65 & 173 &150 &204 &108 & 66 & 74 & 92\\
$m_{\chi ^0 _3} $ & 455. & 219 & 310 &391 &445 &336 &170 &191 &208\\
$m_{\chi ^0 _4} $ & 471. & 263 & 342 &409 &472 &361 &208 &236 &244\\
$m_{\tilde{e}_L}$ & 328. & 124 & 211 &426 &472 &310 & 90 & 94 &109\\
$m_{\tilde{u}_L}$ & 700. & 283 & 570 &590 &664 &449 &251 &275 &319\\
$m_{\tilde{d}_R}$ & 676. & 276 & 550 &577 &638 &441 &246 &268 &310\\
$m_{\tilde{g}}  $ & 718. & 292 & 610 &483 &706 &371 &280 &304 &350\\
\hline
$C_{\tilde{\chi}^0}\times10^{-3}$ & 3.3 & 0.023 & 0.46 & 5.9 &
                14 & 1.4 & 0.0068 & 0.0089 &0.019\\
$C_{\tilde{g}}\times10^{-3}$       & 14 & 1.6 & 54 & 56 & 110 & 13 & 0.97 & 
                                   1.5 & 3.1\\
\hline
$C_{\tilde{\chi}^0,\tilde{g}}\times10^{-3}$ & 3.2 & 0.023 & 0.45 & 
  5.3 & 12 & 1.3 & 0.0068 & 0.0089 & 0.019\\ 
\bottomline
\end{tabular}
\end{table}
From Table 6 we see that  there is quite spread in the quantities
$C_{\tilde{\chi}^0},C_{\tilde{g}}$ 
and $C_{\tilde{\chi}^0,\tilde{g}}$, depending
on the SUSY parameter space. We will see that this is the largest uncertainty
in estimating the SUSY contribution to $0\nu \beta \beta$ decay.  In all
of these cases the intermediate selectron-neutralino mechanism appears to be 
the most dominant.The most 
favorable situation occurs in the case $\#$ 7 of Table 6. And this what we
will consider in extracting the limits on $\lambda^{\prime}_{111}$
Combining the above values of the couplings $\alpha_{k\pi}$, k=1,2,
with the corresponding nuclear matrix elements of of Faessler et al 
\cite{FKS98a}) (F) and the two nucleon ME of Wodecki et al \cite{WKS99} (W)
we obtain the limits listed as Pr in Table 7.
\begin{table}[t]
\footnotesize\rm
\caption{The limits for $\eta_{SUSY}$ and $\lambda^{\prime}_{111}$ obtained :
 a) For the pion mechanism
using the values of $\alpha_{1\pi}$ and $\alpha_{2\pi}$ computed in this 
work and the nuclear ME of Faessler et al (F) 
b) Using the nuclear ME of
the two nucleon mode of Ref. Wodecki et al (W). In extracting the values of
$\lambda^{\prime}_{111}$ we used the SUSY data of $\#$7 of Table 6.
The experimental lifetimes employed are those of Table 4.}
\label{table.7}
\begin{tabular}{lrrrrrr}
\topline
 &\multicolumn{4}{c}{ Pion~mode} & 
  \multicolumn{2}{c}{Only~nucleons} \\ 
\hline
(A,Z) & $\eta_{SUSY}(Pr)$ & $\lambda^{\prime}_{111}(Pr)$ & 
             $\eta_{SUSY}(F)$ & $\lambda^{\prime}_{111}(F)$ &
             $\eta_{SUSY}(Pr)$ & $\lambda^{\prime}_{111}(Pr)$\\
$^{76}Ge$  & $8.4\times10^{-9}$ & $6.0\times10^{-4}$ & $5.5\times10^{-9}$ & 
             $4.8\times10^{-4}$ & $2.6\times10^{-8}$ & $1.1\times10^{-3}$\\
$^{100}Mo$ & $3.2\times10^{-8}$ & $1.8\times10^{-3}$ & $2.4\times10^{-8}$ & 
             $1.1\times10^{-3}$ & $1.1\times10^{-7}$ & $2.2\times10^{-3}$\\
$^{116}Cd$ & $7.6\times10^{-8}$ & $1.8\times10^{-3}$ & $5.4\times10^{-8}$ & 
             $1.5\times10^{-3}$ & $2.6\times10^{-7}$ & $3.3\times10^{-3}$\\
$^{128}Te$ & $1.6\times10^{-8}$ & $8.1\times10^{-4}$ & $1.1\times10^{-8}$ & 
             $6.8\times10^{-4}$ & $5.6\times10^{-8}$ & $1.6\times10^{-3}$\\
$^{130}Te$ & $1.0\times10^{-7}$ & $1.8\times10^{-3}$ & $5.5\times10^{-8}$ & 
             $1.5\times10^{-3}$ & $1.3\times10^{-7}$ & $6.5\times10^{-3}$\\
$^{136}Xe$ & $2.4\times10^{-8}$ & $9.4\times10^{-4}$ & $1.7\times10^{-8}$ & 
             $7.8\times10^{-4}$ & $8.7\times10^{-7}$ & $2.2\times10^{-3}$\\
$^{150}Nd$ & $7.3\times10^{-8}$ & $2.3\times10^{-3}$ & $5.2\times10^{-8}$ & 
             $1.4\times10^{-3}$ & $2.4\times10^{-7}$ & $3.0\times10^{-3}$\\
\bottomline
\end{tabular}
\end{table}
 
Thus the most stringent limit is obtained from the $^{76}Ge$ data and is
\beq
\lambda^{\prime}_{111} \le 6.0\times 10^{-4} ~~(for~ case~\#7~)
\label{eq:6.15}   
\eeq
The above quantities are assumed positive. If not, the absolute value is
understood.

\begin{table}[t]
\footnotesize\rm
\caption{Summary of the results presented in this work.} 
\label{table.8}
\begin{tabular}{lrrrrrr}
\topline
 & $\langle m_{\nu}\rangle$ & $\langle \lambda \rangle$ & 
   $\langle \eta \rangle$ & $\langle \eta _N \rangle$ &
   $\langle \eta _{SUSY} \rangle$ & $\lambda^{\prime}_{111}$\\
\hline
 (A,Z) & eV & $10\times^{-6}$ & $10\times^{-8}$ & $10\times^{-8}$ &
              $10\times^{-8}$ & $10\times^{-4}$ \\
\hline
  & Pr & P1 & P1 & Pr & Pr & Pr \\
\hline
$^{76}Ge$  & 0.27 & 0.56 & 0.32 & 0.44 & 0.31 & 4.0 \\
$^{100}Mo$ & 2.9 & 26 & 8.8 & 6.2 & 3.2 & 18 \\ 
$^{116}Cd$ & 5.9 & 37 & 26 & 11 & 7.6 & 18 \\ 
$^{128}Te$ & 1.8 & 5.6 & 1.3 & 2.9 & 1.6 & 8.1 \\ 
$^{130}Te$ & 13 & 7.6 & 5.2 & 20 & 10& 18 \\ 
$^{136}Xe$ & 49 & 2.1&1.4 & 4 5 & 2.4 & 9.4 \\ 
$^{150}Nd$ & 8.5 & 5.6 & 5.3 & 16 & 7.3 & 23 \\ 
\bottomline
\end{tabular}
\end{table}

\section{Conclusions}
 We have seen that $0\nu \beta \beta $ decay pops up in almost any
fashionable particle model. Thus it can  set useful limits not only
on the light neutrino mass (\ref{vio.2}), but in addition on other lepton
violating parameters like $<\eta _N >$ of (\ref{vio.2}) or the parameters
$\lambda $ and $\eta $ (see sect. 2.2). Finally we mention again the limit 
extracted on the R-parity violating parameter (\ref{eq:6.15}).   
A set of limits, for our choice of nuclear
matrix elements, derived from the various nuclear targets is given in Table 8.
For $^{76}Ge$ our results are different from the ones given above, since we have
used here the unpublished new limit of the Heidelberg-
Moscow experiment $T_{1/2} \ge 5.7 \times 10^{25}$ yr.

We see that limits are quite stringent, but they, of course, have uncertainties
in them. They come from nuclear physics, especially for the short ranged
 operators or from particle physics, as, e.g., in the case of supersymmetry.
It is however evident that in the extraction of $\lambda^{\prime}_{111}$ the
main uncertainty comes from the parameters of supersymmetry. After all
$\lambda^{\prime}_{111}$ depends on the inverse fourth root of the lifetime
and the inverse square root of the nuclear matrix elements. On the other hand
it depends in the second power of the masses of the mediating SUSY scalars.
This is not true of the models considered here but it is found in other
calculations as well \cite{WKS99}.

It is clear that during the last year the interest of most people is being 
focused on the light neutrino mass mechanism. This due to the experimental 
indications for neutrino 
oscillations  of solar (Homestake \cite{cle95}, Kamiokande \cite{hir91}, 
Gallex \cite{gal95} and SAGE \cite{abd94}), atmospheric 
(Kamiokande \cite{fuk94}, IMB \cite{bec95} and 
Soudan \cite{goo96}, Super-Kamiokande \cite{Fuk98})
and terrestrial (LSND experiment \cite{ath95}) experiments.

One can use the constraints imposed by the results  of neutrino 
oscillation experiments on $<\Delta m^2_\nu >$. These experiments, of
course, cannot predict the scale of the masses or the Majorana phases.
The predictions differ 
from each other due the different input and structure of the 
neutrino  mixing matrix and assumptions.
 Bilenky et al \cite{biln98} and others \cite{BFK98} have shown
that under quite reasonable
assumptions in a general scheme with three light Majorana neutrinos and 
mass hierarchy  $|<m_\nu >|$ is smaller than $10^{-2}$ eV. 
In another study  outlined in Ref.\cite{BS98} the authors end up
with $|<m_\nu >| \approx 0.14$ eV. Thus one can see that, the
current limit on $<m_\nu >$ in (\ref{vio.2}) is quite a bit higher than
the neutrino oscillation data.

There is 
a new experimental proposal for measurement of the $0\nu\beta\beta$-decay
of $^{76}Ge$, which intents to use 1 ton (in an extended version 10 tons)
of enriched $^{76}Ge$ and to reach the the half-life limit
$T^{0\nu -exp}_{1/2} \geq 5.8\times 10^{27}$ and  
$T^{0\nu -exp}_{1/2} \geq 6.4\times 10^{28}$ after one and 10 years
of measurements, respectively. From these half-life values one can 
deduce [see Eq. (\ref{eq:1.4}) and Table \ref{table.2}] the possible
future limits on the effective light neutrino mass 
$2.7\times 10^{-2}$ eV and $8.1\times 10^{-3}$ eV, respectively. 
 From the comparison with limits advocated by the neutrino oscillation
phenomenology we conclude that GENIUS experiment \cite {hdmo97,HK97} would
be able to measure this lepton number violating process, provided, of course, 
that the neutrinos are Majorana particles.

We must emphasize that the plethora of other $0\nu\beta\beta$-decay
mechanisms  predicted by GUT's and SUSY do not diminish the importance of
this reaction in settling the outstanding neutrino properties. One can show
that the presence of these exotic mechanisms implies that the neutrinos are
massive Majorana particles, even if the mass mechanism is not the 
dominant one \cite{SVa82,kov97}.

 Thus one can say with certainty that the 
experimental detection of the $0\nu\beta\beta$-decay process would be
a  major achievement with important implications 
on the field of particle and nuclear physics
as well as on cosmology.

\section*{Acknowledgments}

I  would like to express my appreciation to the
Humboldt Foundation for their award and my thanks to the 
Institute of Theoretical Physics at the University of T\" ubingen
for its hospitality.  I am also indebted to F. Simkovic and O. Haug for
their help in the preparation of the manuscript.


\end{document}